\documentclass[11pt]{article}

\begin{document}

\title{Antisymmetric Tensor Fields, 4-Vector Fields, Indefinite Metrics and Normalization\thanks{Talk given at  the {\it VII Mexican School on Gravitation and Mathematical Physics "Relativistic Astrophysics and Numerical Relativity"}, November 26 -- December 1, 2006,
Playa del Carmen, QR, M\'exico; and the 10th Workshop ``{\it What comes beyond the Standard Model?}", July 17-27, 2007, Bled, Slovenia.}}

\author{Valeri V. Dvoeglazov\\
Universidad de Zacatecas, Apartado Postal 636,
Suc. UAZ\\Zacatecas 98062, Zac., M\'exico\\
E-mail: valeri@planck.reduaz.mx,\\
URL: http://planck.reduaz.mx/\~\,valeri/
}

\maketitle


\begin{abstract}
On the basis of our recent modifications of the Dirac formalism we 
generalize the Bargmann-Wigner formalism for higher spins to be compatible 
with other formalisms for bosons. Relations with dual electrodynamics, with 
the Ogievetskii-Polubarinov notoph and the Weinberg 2(2J+1) theory are 
found. Next, we introduce the dual analogues of the Riemann tensor and 
derive corresponding dynamical equations in the Minkowski space. Relations 
with the Marques-Spehler chiral gravity theory are discussed. The problem of 
indefinite metrics, particularly, in quantization of 4-vector fields is 
clarified.
\end{abstract}

\section{Introduction}

The general scheme for derivation of higher-spin equations
was given in~\cite{bw-hs}. A field of rest mass $m$ and spin $j \geq {1\over
2}$ is represented by a completely symmetric multispinor of rank $2j$.
The particular cases $j=1$ and $j={3\over 2}$ were given in the
textbooks, e.~g., ref.~\cite{Lurie}. Generalized equations for higher spins can be derived
from the first principles on using some modifications of the Bargmann-Wigner formalism.
The generalizations of the equations in the $(1/2,0)\oplus (0,1/2)$ representation are 
well known. The Tokuoka-SenGupta-Fushchich formalism and the Barut formalism are based on 
the equations presented in refs.~\cite{g1,g2,g3,g3a,Fush,gts,barut,Wilson,Dvo,afd-dvo}.

\section{Generalized Spin-1 Case}

We begin with
\begin{eqnarray}
\left [ i\gamma_\mu \partial_\mu + a -b \partial^2 + \gamma_5 (c- d\partial^2 )
\right ]_{\alpha\beta} \Psi_{\beta\gamma} &=&0\,,\\
\left [ i\gamma_\mu
\partial_\mu + a -b \partial^2 - \gamma_5 (c- d\partial^2 ) \right ]_{\alpha\beta}
\Psi_{\gamma\beta} &=&0\,,
\end{eqnarray}
$\partial^2$ is the d'Alembertian.
Thus, we obtain the Proca-like equations:
\begin{eqnarray} &&\partial_\nu A_\lambda - \partial_\lambda A_\nu - 2(a
+b \partial_\mu \partial_\mu ) F_{\nu \lambda} =0\,,\\ &&\partial_\mu
F_{\mu \lambda} = {1\over 2} (a +b \partial_\mu \partial_\mu) A_\lambda +
{1\over 2} (c+ d \partial_\mu \partial_\mu) \tilde A_\lambda\,,
\end{eqnarray}
$\tilde A_\lambda$ is the axial-vector potential (analogous to that
used in the Duffin-Kemmer set of equations for $J=0$). Additional constraints are:
\begin{eqnarray}
&&i\partial_\lambda A_\lambda + ( c+d\partial_\mu \partial_\mu) \tilde \phi
=0\,,\\
&&\epsilon_{\mu\lambda\kappa \tau} \partial_\mu F_{\lambda\kappa } =0\,,
( c+ d \partial_\mu \partial_\mu ) \phi =0\,.
\end{eqnarray}

The spin-0 Duffin-Kemmer equations are:
\begin{eqnarray}
&&(a+b \partial_\mu \partial_\mu) \phi = 0\,, 
i\partial_\mu \tilde A_\mu  - (a+b\partial_\mu \partial_\mu) \tilde
\phi =0\,,\\
&&(a+b\, \partial_\mu \partial_\mu) \tilde A_\nu + (c+d\,\partial_\mu
\partial_\mu) A_\nu + i (\partial_\nu \tilde \phi) =0\,.
\end{eqnarray}
The additional constraints are:
\begin{equation}
\partial_\mu \phi =0\,,
\partial_\nu \tilde A_\lambda - \partial_\lambda \tilde
A_\nu +2 (c+d\partial_\mu \partial_\mu ) F_{\nu \lambda} = 0\,.
\end{equation}
In such a way the spin states are {\it mixed} through the 4-vector potentials.
After elimination of the 4-vector potentials we obtain
the equation for the AST field of the second rank:
\begin{eqnarray}
\lefteqn{\left [ \partial_\mu \partial_\nu F_{\nu\lambda} - \partial_\lambda
\partial_\nu F_{\nu\mu}\right ]   +}\nonumber\\
&&+\left [ (c^2 - a^2) - 2(ab-cd)
\partial_\mu\partial_\mu  + (d^2 -b^2)
(\partial_\mu\partial_\mu)^2 \right ] F_{\mu\lambda} = 0\,,
\end{eqnarray}
which should be compared with our
previous equations which follow from the Weinberg-like formulation~\cite{Weinberg,dvo-rmf,dvo-hpa}.
Just put:
\begin{eqnarray}
c^2 - a^2 \Rightarrow {-Bm^2 \over 2}\,,&\qquad& c^2 - a^2 \Rightarrow
+{Bm^2 \over 2}\,,\\
-2(ab-cd) \Rightarrow {A-1\over 2}\,,&\qquad&
+2(ab-cd) \Rightarrow {A+1\over 2}\,,\\
b=\pm d\,.&\qquad&
\end{eqnarray}
Of course, these sets of algebraic equations have solutions in terms $A$
and $B$. We found them and restored the equations.
The parity violation and the spin mixing are {\it intrinsic} possibilities
of the Proca-like theories.

In fact, there  are several modifications of the BW formalism. One can propose the following set:
\begin{eqnarray}
\left [ i\gamma_\mu \partial_\mu + \epsilon_1 m_1 +\epsilon_2 m_2 \gamma_5
\right ]_{\alpha\beta} \Psi_{\beta\gamma} &=&0\,,\\
\left [ i\gamma_\mu
\partial_\mu + \epsilon_3 m_1 +\epsilon_4 m_2 \gamma_5 \right ]_{\alpha\beta}
\Psi_{\gamma\beta} &=&0\,,
\end{eqnarray}
where $\epsilon_i = i\partial_t/E$ are the sign operators. So, at first sight, we have 16
possible combinations for the AST fields. We first come to
\begin{eqnarray}
&&\left [ i\gamma_\mu \partial_\mu + m_1 A_1 + m_2 A_2\gamma_5
\right ]_{\alpha\beta} \left \{ (\gamma_\lambda R)_{\beta\gamma} A_\lambda
+ (\sigma_{\lambda\kappa } R)_{\beta\gamma} F_{\lambda\kappa }\right
\}+\nonumber\\ &+&\left [ m_1 B_1 +m_2 B_2 \gamma_5 \right ]_{\alpha\beta} \left \{
R_{\beta\gamma}\varphi + (\gamma_5 R)_{\beta\gamma} \tilde \phi +(\gamma_5
\gamma_\lambda R)_{\beta\gamma} \tilde A_\lambda\right \}=0\,,\\
&&\left [
i\gamma_\mu \partial_\mu + m_1 A_1 + m_2 A_2\gamma_5 \right
]_{\gamma\beta} \left \{ (\gamma_\lambda R)_{\alpha\beta} A_\lambda +
(\sigma_{\lambda\kappa } R)_{\alpha\beta} F_{\lambda\kappa }\right \}-\nonumber\\
&-&\left [ m_1 B_1 +m_2 B_2 \gamma_5 \right ]_{\alpha\beta} \left \{
R_{\alpha\beta}\varphi +(\gamma_5 R)_{\alpha\beta} \tilde \phi +(\gamma_5
\gamma_\lambda R)_{\alpha\beta} \tilde A_\lambda\right \}=0\,,
\end{eqnarray}
where $A_1 = {\epsilon_1 +\epsilon_3 \over 2}$,
$A_2 = {\epsilon_2 +\epsilon_4 \over 2}$,
$B_1 = {\epsilon_1 -\epsilon_3 \over 2}$,
and
$B_2 = {\epsilon_2 -\epsilon_4 \over 2}$.
Thus, for spin 1 we have
\begin{eqnarray} &&\partial_\mu A_\lambda - \partial_\lambda A_\mu + 2m_1 A_1 F_{\mu \lambda}
+im_2 A_2 \epsilon_{\alpha\beta\mu\lambda} F_{\alpha\beta} =0\,,\\
&&\partial_\lambda
F_{\kappa \lambda} - {m_1\over 2} A_1 A_\kappa -{m_2\over 2} B_2 \tilde
A_\kappa=0\,,
\end{eqnarray}  with constraints
\begin{eqnarray}
&&-i\partial_\mu A_\mu + 2m_1 B_1 \phi +2m_2 B_2 \tilde \phi=0\,,\\
&&i\epsilon_{\mu\nu\kappa\lambda} \partial_\mu F_{\nu\kappa}
-m_2 A_2 A_\lambda -m_1 B_1 \tilde A_\lambda =0\,,\\
&&m_1 B_1 \tilde \phi +m_2 B_2 \phi =0\,.
\end{eqnarray}
If we remove $A_\lambda$ and $\tilde A_\lambda$ from this set,
we come to the final results for the AST field.
Actually, we have twelve equations, see~\cite{dvo-wig}. One can
go even further. One can use the Barut equations for the BW input. So, we
can get $16\times 16$ combinations (depending on the eigenvalues of the
corresponding sign operators), and we have different eigenvalues of masses 
due to $\partial_\mu^2 = \kappa m^2$.

Why do I think that the shown arbitrarieness
of equations for the AST fields is related to 1) spin basis rotations; 2)
the choice of normalization? (see ref.~\cite{dv-ps})  In the common-used basis  three
4-potentials have parity eigenvalues $-1$ and one time-like (or spin-0
state), $+1$; the fields ${\bf E}$ and ${\bf B}$ have also definite parity
properties in this basis.  If we transfer to other  basis, e.g., to the
helicity basis~\cite{Ber} we can see that the 4-vector potentials and
the corresponding fields are superpositions of a vector and an
axial-vector~\cite{Grei}.  Of course, they can be expanded in the fields in the
``old" basis.

The detailed discussion of the generalized spin-1 case (as well as the problems related to normalization, indefinite metric and 4-vector fields) can be found in refs.~\cite{dvo-wig,dv-ps, dvo-new}.

\section{Generalized Spin-2 Case}

 The spin-2 case can also be of some
interest because it is generally believed that the essential features of
the gravitational field are  obtained from transverse components of the
$(2,0)\oplus (0,2)$  representation of the Lorentz group. Nevertheless,
questions of the redandant components of the higher-spin relativistic
equations are not yet understood in detail~\cite{Kirch}.

We begin with the commonly-accepted procedure
for the derivation  of higher-spin equations below.
We begin with the equations for the 4-rank symmetric spinor:
\begin{eqnarray}
&&\left [ i\gamma^\mu \partial_\mu - m \right ]_{\alpha\alpha^\prime}
\Psi_{\alpha^\prime \beta\gamma\delta} = 0\, ,
\left [ i\gamma^\mu \partial_\mu - m \right ]_{\beta\beta^\prime}
\Psi_{\alpha\beta^\prime \gamma\delta} = 0\, ,\\
&&\left [ i\gamma^\mu \partial_\mu - m \right ]_{\gamma\gamma^\prime}
\Psi_{\alpha\beta\gamma^\prime \delta} = 0\, ,
\left [ i\gamma^\mu \partial_\mu - m \right ]_{\delta\delta^\prime}
\Psi_{\alpha\beta\gamma\delta^\prime} = 0\, .
\end{eqnarray} 
The massless limit (if one needs) should be taken in the end of all
calculations.

We proceed expanding the field function in the complete set of symmetric matrices
(as in the spin-1 case). In the beginning let us use the
first two indices:
\begin{equation} \Psi_{\{\alpha\beta\}\gamma\delta} =
(\gamma_\mu R)_{\alpha\beta} \Psi^\mu_{\gamma\delta}
+(\sigma_{\mu\nu} R)_{\alpha\beta} \Psi^{\mu\nu}_{\gamma\delta}\, .
\end{equation}
We would like to write
the corresponding equations for functions $\Psi^\mu_{\gamma\delta}$
and $\Psi^{\mu\nu}_{\gamma\delta}$ in the form:
\begin{equation}
{2\over m} \partial_\mu \Psi^{\mu\nu}_{\gamma\delta} = -
\Psi^\nu_{\gamma\delta}\, , 
\Psi^{\mu\nu}_{\gamma\delta} = {1\over 2m}
\left [ \partial^\mu \Psi^\nu_{\gamma\delta} - \partial^\nu
\Psi^\mu_{\gamma\delta} \right ]\, \label{p2}.
\end{equation} 
The constraints $(1/m) \partial_\mu \Psi^\mu_{\gamma\delta} =0$
and $(1/m) \epsilon^{\mu\nu}_{\quad\alpha\beta}\, \partial_\mu
\Psi^{\alpha\beta}_{\gamma\delta} = 0$ can be regarded as the consequence of
Eqs.  (\ref{p2}).
Next, we present the vector-spinor and tensor-spinor functions as
\begin{eqnarray}
&&\Psi^\mu_{\{\gamma\delta\}} = (\gamma^\kappa R)_{\gamma\delta}
G_{\kappa}^{\quad \mu} +(\sigma^{\kappa\tau} R )_{\gamma\delta}
F_{\kappa\tau}^{\quad \mu} \, ,\\
&&\Psi^{\mu\nu}_{\{\gamma\delta\}} = (\gamma^\kappa R)_{\gamma\delta}
T_{\kappa}^{\quad \mu\nu} +(\sigma^{\kappa\tau} R )_{\gamma\delta}
R_{\kappa\tau}^{\quad \mu\nu} \, ,
\end{eqnarray}
i.~e.,  using the symmetric matrix coefficients in indices $\gamma$ and
$\delta$. Hence, the total function is
\begin{eqnarray}
\lefteqn{\Psi_{\{\alpha\beta\}\{\gamma\delta\}}
= (\gamma_\mu R)_{\alpha\beta} (\gamma^\kappa R)_{\gamma\delta}
G_\kappa^{\quad \mu} + (\gamma_\mu R)_{\alpha\beta} (\sigma^{\kappa\tau}
R)_{\gamma\delta} F_{\kappa\tau}^{\quad \mu} + } \nonumber\\
&+& (\sigma_{\mu\nu} R)_{\alpha\beta} (\gamma^\kappa R)_{\gamma\delta}
T_\kappa^{\quad \mu\nu} + (\sigma_{\mu\nu} R)_{\alpha\beta}
(\sigma^{\kappa\tau} R)_{\gamma\delta} R_{\kappa\tau}^{\quad\mu\nu} \, ;
\end{eqnarray}
and the resulting tensor equations are:
\begin{eqnarray}
&&{2\over m} \partial_\mu T_\kappa^{\quad \mu\nu} =
-G_{\kappa}^{\quad\nu}\, ,
{2\over m} \partial_\mu R_{\kappa\tau}^{\quad \mu\nu} =
-F_{\kappa\tau}^{\quad\nu}\, ,\\
&& T_{\kappa}^{\quad \mu\nu} = {1\over 2m} \left [
\partial^\mu G_{\kappa}^{\quad\nu}
- \partial^\nu G_{\kappa}^{\quad \mu} \right ] \, ,\\
&& R_{\kappa\tau}^{\quad \mu\nu} = {1\over 2m} \left [
\partial^\mu F_{\kappa\tau}^{\quad\nu}
- \partial^\nu F_{\kappa\tau}^{\quad \mu} \right ] \, .
\end{eqnarray}
The constraints are re-written to
\begin{eqnarray}
&&{1\over m} \partial_\mu G_\kappa^{\quad\mu} = 0\, ,\quad
{1\over m} \partial_\mu F_{\kappa\tau}^{\quad\mu} =0\, ,\\
&& {1\over m} \epsilon_{\alpha\beta\nu\mu} \partial^\alpha
T_\kappa^{\quad\beta\nu} = 0\, ,\quad
{1\over m} \epsilon_{\alpha\beta\nu\mu} \partial^\alpha
R_{\kappa\tau}^{\quad\beta\nu} = 0\, .
\end{eqnarray}
However, we need to make symmetrization over these two sets
of indices $\{ \alpha\beta \}$ and $\{\gamma\delta \}$. The total
symmetry can be ensured if one contracts the function $\Psi_{\{\alpha\beta
\} \{\gamma \delta \}}$ with {\it antisymmetric} matrices
$R^{-1}_{\beta\gamma}$, $(R^{-1} \gamma^5 )_{\beta\gamma}$ \linebreak
and
$(R^{-1} \gamma^5 \gamma^\lambda )_{\beta\gamma}$ and equate
all these contractions to zero (similar to the $j=3/2$ case
considered in ref.~\cite[p. 44]{Lurie}. We obtain
additional constraints on the tensor field functions:
\begin{eqnarray}
&& G_\mu^{\quad\mu}=0\, , \quad G_{[\kappa \, \mu ]}  = 0\, , \quad
G^{\kappa\mu} = {1\over 2} g^{\kappa\mu} G_\nu^{\quad\nu}\, ,
\label{b1}\\
&&F_{\kappa\mu}^{\quad\mu} = F_{\mu\kappa}^{\quad\mu} = 0\, , \quad
\epsilon^{\kappa\tau\mu\nu} F_{\kappa\tau,\mu} = 0\, ,\\
&& T^{\mu}_{\quad\mu\kappa} =
T^{\mu}_{\quad\kappa\mu} = 0\, ,\quad
\epsilon^{\kappa\tau\mu\nu} T_{\kappa,\tau\mu} = 0\, ,\\
&& F^{\kappa\tau,\mu} = T^{\mu,\kappa\tau}\, ,\quad
\epsilon^{\kappa\tau\mu\lambda} (F_{\kappa\tau,\mu} +
T_{\kappa,\tau\mu})=0\, ,\\
&& R_{\kappa\nu}^{\quad \mu\nu}
= R_{\nu\kappa}^{\quad  \mu\nu} = R_{\kappa\nu}^{\quad\nu\mu}
= R_{\nu\kappa}^{\quad\nu\mu}
= R_{\mu\nu}^{\quad  \mu\nu} = 0\, , \\
&& \epsilon^{\mu\nu\alpha\beta} (g_{\beta\kappa} R_{\mu\tau,
\nu\alpha} - g_{\beta\tau} R_{\nu\alpha,\mu\kappa} ) = 0\, \quad
\epsilon^{\kappa\tau\mu\nu} R_{\kappa\tau,\mu\nu} = 0\, .\label{f1}
\end{eqnarray} 
Thus, we  encountered with
the known difficulty of the theory for spin-2 particles in
the Minkowski space.
We explicitly showed that all field functions become to be equal to zero.
Such a situation cannot be considered as a satisfactory one (because it
does not give us any physical information) and can be corrected in several
ways.

We shall modify the formalism~\cite{dv-ps}. The field function is now presented as
\begin{equation}
\Psi_{\{\alpha\beta\}\gamma\delta} =
\alpha_1 (\gamma_\mu R)_{\alpha\beta} \Psi^\mu_{\gamma\delta} +
\alpha_2 (\sigma_{\mu\nu} R)_{\alpha\beta} \Psi^{\mu\nu}_{\gamma\delta}
+\alpha_3 (\gamma^5 \sigma_{\mu\nu} R)_{\alpha\beta}
\widetilde \Psi^{\mu\nu}_{\gamma\delta}\, ,
\end{equation}
with
\begin{eqnarray}
&&\Psi^\mu_{\{\gamma\delta\}} = \beta_1 (\gamma^\kappa R)_{\gamma\delta}
G_\kappa^{\,\,\mu} + \beta_2 (\sigma^{\kappa\tau} R)_{\gamma\delta}
F_{\kappa\tau}^{\,\,\mu} +\beta_3 (\gamma^5 \sigma^{\kappa\tau}
R)_{\gamma\delta} \widetilde F_{\kappa\tau}^{\quad\mu} ,\\
&&\Psi^{\mu\nu}_{\{\gamma\delta\}} =\beta_4 (\gamma^\kappa
R)_{\gamma\delta} T_\kappa^{\,\,\mu\nu} + \beta_5 (\sigma^{\kappa\tau}
R)_{\gamma\delta} R_{\kappa\tau}^{\,\,\mu\nu} +\beta_6 (\gamma^5
\sigma^{\kappa\tau} R)_{\gamma\delta}
\widetilde R_{\kappa\tau}^{\,\,\mu\nu},\\
&&\widetilde \Psi^{\mu\nu}_{\{\gamma\delta\}} =\beta_7 (\gamma^\kappa
R)_{\gamma\delta} \widetilde T_\kappa^{\,\,\mu\nu} + \beta_8
(\sigma^{\kappa\tau} R)_{\gamma\delta}
\widetilde D_{\kappa\tau}^{\,\,\mu\nu}
+\beta_9 (\gamma^5 \sigma^{\kappa\tau} R)_{\gamma\delta}
D_{\kappa\tau}^{\,\,\mu\nu} .
\end{eqnarray}
Hence, the function $\Psi_{\{\alpha\beta\}\{\gamma\delta\}}$
can be expressed as a sum of nine terms:
\begin{eqnarray}
&&\Psi_{\{\alpha\beta\}\{\gamma\delta\}} =
\alpha_1 \beta_1 (\gamma_\mu R)_{\alpha\beta} (\gamma^\kappa
R)_{\gamma\delta} G_\kappa^{\quad\mu} +\alpha_1 \beta_2
(\gamma_\mu R)_{\alpha\beta} (\sigma^{\kappa\tau} R)_{\gamma\delta}
F_{\kappa\tau}^{\quad\mu} + \nonumber\\
&+&\alpha_1 \beta_3 (\gamma_\mu R)_{\alpha\beta}
(\gamma^5 \sigma^{\kappa\tau} R)_{\gamma\delta} \widetilde
F_{\kappa\tau}^{\quad\mu} +
+ \alpha_2 \beta_4 (\sigma_{\mu\nu}
R)_{\alpha\beta} (\gamma^\kappa R)_{\gamma\delta} T_\kappa^{\quad\mu\nu}
+\nonumber\\
&+&\alpha_2 \beta_5 (\sigma_{\mu\nu} R)_{\alpha\beta} (\sigma^{\kappa\tau}
R)_{\gamma\delta} R_{\kappa\tau}^{\quad \mu\nu}
+ \alpha_2
\beta_6 (\sigma_{\mu\nu} R)_{\alpha\beta} (\gamma^5 \sigma^{\kappa\tau}
R)_{\gamma\delta} \widetilde R_{\kappa\tau}^{\quad\mu\nu} +\nonumber\\
&+&\alpha_3 \beta_7 (\gamma^5 \sigma_{\mu\nu} R)_{\alpha\beta}
(\gamma^\kappa R)_{\gamma\delta} \widetilde
T_\kappa^{\quad\mu\nu}+
\alpha_3 \beta_8 (\gamma^5
\sigma_{\mu\nu} R)_{\alpha\beta} (\sigma^{\kappa\tau} R)_{\gamma\delta}
\widetilde D_{\kappa\tau}^{\quad\mu\nu} +\nonumber\\
&+&\alpha_3 \beta_9
(\gamma^5 \sigma_{\mu\nu} R)_{\alpha\beta} (\gamma^5 \sigma^{\kappa\tau}
R)_{\gamma\delta} D_{\kappa\tau}^{\quad \mu\nu}\, .
\label{ffn1}
\end{eqnarray}
The corresponding dynamical
equations are given by the set of equations
\begin{eqnarray}
&& {2\alpha_2
\beta_4 \over m} \partial_\nu T_\kappa^{\quad\mu\nu} +{i\alpha_3
\beta_7 \over m} \epsilon^{\mu\nu\alpha\beta} \partial_\nu
\widetilde T_{\kappa,\alpha\beta} = \alpha_1 \beta_1
G_\kappa^{\quad\mu}\,; \label{b}\\
&&{2\alpha_2 \beta_5 \over m} \partial_\nu
R_{\kappa\tau}^{\quad\mu\nu} +{i\alpha_2 \beta_6 \over m}
\epsilon_{\alpha\beta\kappa\tau} \partial_\nu \widetilde R^{\alpha\beta,
\mu\nu} +{i\alpha_3 \beta_8 \over m}
\epsilon^{\mu\nu\alpha\beta}\partial_\nu \widetilde
D_{\kappa\tau,\alpha\beta} - \nonumber\\
&-&{\alpha_3 \beta_9 \over 2}
\epsilon^{\mu\nu\alpha\beta} \epsilon_{\lambda\delta\kappa\tau}
D^{\lambda\delta}_{\quad \alpha\beta} = \alpha_1 \beta_2
F_{\kappa\tau}^{\quad\mu} + {i\alpha_1 \beta_3 \over 2}
\epsilon_{\alpha\beta\kappa\tau} \widetilde F^{\alpha\beta,\mu}\,; \\
&& 2\alpha_2 \beta_4 T_\kappa^{\quad\mu\nu} +i\alpha_3 \beta_7
\epsilon^{\alpha\beta\mu\nu} \widetilde T_{\kappa,\alpha\beta}
=  {\alpha_1 \beta_1 \over m} (\partial^\mu G_\kappa^{\quad \nu}
- \partial^\nu G_\kappa^{\quad\mu})\,; \\
&& 2\alpha_2 \beta_5 R_{\kappa\tau}^{\quad\mu\nu} +i\alpha_3 \beta_8
\epsilon^{\alpha\beta\mu\nu} \widetilde D_{\kappa\tau,\alpha\beta}
+i\alpha_2 \beta_6 \epsilon_{\alpha\beta\kappa\tau} \widetilde
R^{\alpha\beta,\mu\nu} -\nonumber\\
&-& {\alpha_3 \beta_9\over 2} \epsilon^{\alpha\beta\mu\nu}
\epsilon_{\lambda\delta\kappa\tau} D^{\lambda\delta}_{\quad \alpha\beta}
=  {\alpha_1 \beta_2 \over m} (\partial^\mu F_{\kappa\tau}^{\quad \nu}
-\partial^\nu F_{\kappa\tau}^{\quad\mu} ) + \nonumber\\
&+& {i\alpha_1 \beta_3 \over 2m}
\epsilon_{\alpha\beta\kappa\tau} (\partial^\mu \widetilde
F^{\alpha\beta,\nu} - \partial^\nu \widetilde F^{\alpha\beta,\mu} )\, .
\label{f}
\end{eqnarray}
The essential constraints are:
\begin{eqnarray}
&&\alpha_1 \beta_1 G^\mu_{\quad\mu} = 0\, ,\quad \alpha_1
\beta_1 G_{[\kappa\mu]} = 0;  
2i\alpha_1 \beta_2 F_{\alpha\mu}^{\quad\mu} +
\alpha_1 \beta_3
\epsilon^{\kappa\tau\mu}_{\quad\alpha} \widetilde F_{\kappa\tau,\mu} =
0;\\
&&2i\alpha_1 \beta_3 \widetilde F_{\alpha\mu}^{\quad\mu}
+ \alpha_1 \beta_2
\epsilon^{\kappa\tau\mu}_{\quad\alpha} F_{\kappa\tau,\mu} = 0\, ;
2i\alpha_2 \beta_4 T^{\mu}_{\quad\mu\alpha} -
 \alpha_3 \beta_{7}
\epsilon^{\kappa\tau\mu}_{\quad\alpha} \widetilde T_{\kappa,\tau\mu}
= 0;\\
&& 2i\alpha_3 \beta_{7} \widetilde
T^{\mu}_{\quad\mu\alpha} -
\alpha_2 \beta_4 \epsilon^{\kappa\tau\mu}_{\quad\alpha}
T_{\kappa,\tau\mu} = 0;\\
&& i\epsilon^{\mu\nu\kappa\tau} \left [ \alpha_2 \beta_6 \widetilde
R_{\kappa\tau,\mu\nu} + \alpha_3 \beta_{8} \widetilde
D_{\kappa\tau,\mu\nu} \right ] + 2\alpha_2 \beta_5
R^{\mu\nu}_{\quad\mu\nu}  + 2\alpha_3
\beta_{9} D^{\mu\nu}_{\quad \mu\nu}  = 0;\\
&& i\epsilon^{\mu\nu\kappa\tau} \left [ \alpha_2 \beta_5 R_{\kappa\tau,
\mu\nu} + \alpha_3 \beta_{9} D_{\kappa\tau, \mu\nu} \right ]
+ 2\alpha_2 \beta_6 \widetilde R^{\mu\nu}_{\quad\mu\nu}
+ 2\alpha_3 \beta_{8} \widetilde D^{\mu\nu}_{\quad\mu\nu}  =0;\\
&& 2i \alpha_2 \beta_5 R_{\beta\mu}^{\,\,\,\mu\alpha} + 2i\alpha_3
\beta_{9} D_{\beta\mu}^{\,\,\,\mu\alpha} + \alpha_2 \beta_6
\epsilon^{\nu\alpha}_{\,\,\,\lambda\beta} \widetilde
R^{\lambda\mu}_{\,\,\,\mu\nu} +\alpha_3 \beta_{8}
\epsilon^{\nu\alpha}_{\,\,\,\lambda\beta} \widetilde
D^{\lambda\mu}_{\,\,\, \mu\nu} = 0;\\
&&2i\alpha_1 \beta_2 F^{\lambda\mu}_{\quad\mu} - 2 i \alpha_2 \beta_4
T_\mu^{\quad\mu\lambda} + \alpha_1 \beta_3 \epsilon^{\kappa\tau\mu\lambda}
\widetilde F_{\kappa\tau,\mu} +\alpha_3 \beta_7
\epsilon^{\kappa\tau\mu\lambda} \widetilde T_{\kappa,\tau\mu} =0\, ;\\
&&2i\alpha_1 \beta_3 \widetilde F^{\lambda\mu}_{\quad\mu} - 2 i \alpha_3
\beta_7 \widetilde T_\mu^{\quad\mu\lambda} + \alpha_1 \beta_2
\epsilon^{\kappa\tau\mu\lambda} F_{\kappa\tau,\mu} +\alpha_2
\beta_4 \epsilon^{\kappa\tau\mu\lambda}  T_{\kappa,\tau\mu} =0;\\
&&\alpha_1 \beta_1 (2G^\lambda_{\quad\alpha} - g^\lambda_{\quad\alpha}
G^\mu_{\quad\mu} ) - 2\alpha_2 \beta_5 (2R^{\lambda\mu}_{\quad\mu\alpha}
+2R_{\alpha\mu}^{\quad\mu\lambda} + g^\lambda_{\quad\alpha}
R^{\mu\nu}_{\quad\mu\nu})+\quad\nonumber\\
&+& 2\alpha_3 \beta_9
(2D^{\lambda\mu}_{\quad\mu\alpha} + 2D_{\alpha\mu}^{\quad\mu\lambda}
+g^\lambda_{\quad\alpha} D^{\mu\nu}_{\quad\mu\nu})+
2i\alpha_3 \beta_8 (\epsilon_{\kappa\alpha}^{\quad\mu\nu}
\widetilde D^{\kappa\lambda}_{\quad\mu\nu} - \nonumber\\
&-&\epsilon^{\kappa\tau\mu\lambda} \widetilde D_{\kappa\tau,\mu\alpha}) 
- 2i\alpha_2 \beta_6 (\epsilon_{\kappa\alpha}^{\quad \mu\nu}
\widetilde R^{\kappa\lambda}_{\quad\mu\nu} -
\epsilon^{\kappa\tau\mu\lambda} \widetilde R_{\kappa\tau,\mu\alpha})
= 0; \\
&& 2\alpha_3 \beta_8 (2\widetilde D^{\lambda\mu}_{\quad\mu\alpha} + 2
\widetilde D_{\alpha\mu}^{\quad\mu\lambda} +g^\lambda_{\quad\alpha}
\widetilde D^{\mu\nu}_{\quad\mu\nu}) - 2\alpha_2 \beta_6 (2\widetilde
R^{\lambda\mu}_{\quad\mu\alpha} +2 \widetilde
R_{\alpha\mu}^{\quad\mu\lambda} \nonumber\\
&+& g^\lambda_{\quad\alpha} \widetilde
R^{\mu\nu}_{\quad\mu\nu}) +
+ 2i\alpha_3 \beta_9 (\epsilon_{\kappa\alpha}^{\quad\mu\nu}
D^{\kappa\lambda}_{\quad\mu\nu}  - \epsilon^{\kappa\tau\mu\lambda}
D_{\kappa\tau,\mu\alpha} ) -\nonumber\\
&-& 2i\alpha_2 \beta_5
(\epsilon_{\kappa\alpha}^{\quad\mu\nu} R^{\kappa\lambda}_{\quad\mu\nu}
- \epsilon^{\kappa\tau\mu\lambda} R_{\kappa\tau,\mu\alpha} ) =0;\\
&&\alpha_1 \beta_2 (F^{\alpha\beta,\lambda} - 2F^{\beta\lambda,\alpha}
+ F^{\beta\mu}_{\quad\mu}\, g^{\lambda\alpha} - F^{\alpha\mu}_{\quad\mu}
\, g^{\lambda\beta} ) - \nonumber\\
&-&\alpha_2 \beta_4 (T^{\lambda,\alpha\beta}
-2T^{\beta,\lambda\alpha} + T_\mu^{\quad\mu\alpha} g^{\lambda\beta} -
T_\mu^{\quad\mu\beta} g^{\lambda\alpha} ) +\nonumber\\
&+&{i\over 2} \alpha_1 \beta_3 (\epsilon^{\kappa\tau\alpha\beta}
\widetilde F_{\kappa\tau}^{\quad\lambda} +
2\epsilon^{\lambda\kappa\alpha\beta} \widetilde F_{\kappa\mu}^{\quad\mu} +
2 \epsilon^{\mu\kappa\alpha\beta} \widetilde F^\lambda_{\quad\kappa,\mu})
-\nonumber\\
&-& {i\over 2} \alpha_3 \beta_7 ( \epsilon^{\mu\nu\alpha\beta} \widetilde
T^{\lambda}_{\quad\mu\nu} +2 \epsilon^{\nu\lambda\alpha\beta} \widetilde
T^\mu_{\quad\mu\nu} +2 \epsilon^{\mu\kappa\alpha\beta} \widetilde
T_{\kappa,\mu}^{\quad\lambda} ) =0.
\end{eqnarray}
They are  the results of contractions of the field function (\ref{ffn1})
with three antisymmetric matrices, as above. Furthermore,
one should recover the relations (\ref{b1}-\ref{f1}) in the particular
case when $\alpha_3 = \beta_3 =\beta_6 = \beta_9 = 0$ and
$\alpha_1 = \alpha_2 = \beta_1 =\beta_2 =\beta_4
=\beta_5 = \beta_7 =\beta_8 =1$.

As a discussion we note that in such a framework we have physical
content because only certain combinations of field functions
would be equal to zero. In general, the fields
$F_{\kappa\tau}^{\quad\mu}$, $\widetilde F_{\kappa\tau}^{\quad\mu}$,
$T_{\kappa}^{\quad\mu\nu}$, $\widetilde T_{\kappa}^{\quad\mu\nu}$, and
$R_{\kappa\tau}^{\quad\mu\nu}$,  $\widetilde
R_{\kappa\tau}^{\quad\mu\nu}$, $D_{\kappa\tau}^{\quad\mu\nu}$, $\widetilde
D_{\kappa\tau}^{\quad\mu\nu}$ can  correspond to different physical states
and the equations above describe some kind of ``oscillations" of one state to another.
Furthermore, from the set of equations (\ref{b}-\ref{f}) one
obtains the {\it second}-order equation for symmetric traceless tensor of
the second rank ($\alpha_1 \neq 0$, $\beta_1 \neq 0$):
\begin{equation} {1\over m^2} \left [\partial_\nu
\partial^\mu G_\kappa^{\quad \nu} - \partial_\nu \partial^\nu
G_\kappa^{\quad\mu} \right ] =  G_\kappa^{\quad \mu}\, .
\end{equation}
After the contraction in indices $\kappa$ and $\mu$ this equation is
reduced to the set
\begin{eqnarray}
&&\partial_\mu G_{\quad\nu}^{\mu} = F_\nu\,  \\
&&{1\over m^2} \partial_\nu F^\nu = 0\, ,
\end{eqnarray}
i.~e.,  to the equations connecting the analogue of the energy-momentum
tensor and the analogue of the 4-vector potential. 
Further investigations may provide additional foundations to
``surprising" similarities of gravitational and electromagnetic
equations in the low-velocity limit, refs.~\cite{Wein2,Jef}.

\section*{Acknowledgements} 

I am grateful to participants of recent conferences for discussions.


\end{document}